\documentclass[12pt,a4paper]{article}
\pdfoutput=1
\usepackage{jheppub,tcolorbox}
\usepackage[utf8]{inputenc}
\usepackage{amsmath,amssymb}
\usepackage{float}
\usepackage{tikz} 
\usetikzlibrary{decorations.markings} 
\usetikzlibrary{decorations.text} 
\usetikzlibrary{positioning} 
\usetikzlibrary{backgrounds} 
\usetikzlibrary{fit} 
\usetikzlibrary{calc} 
\usetikzlibrary{shapes} 

\tikzset{dot/.style={draw,circle,inner sep=.7pt,fill,node
    distance=1cm}} 
\tikzset{dot1/.style={draw,circle,inner sep=.7pt,fill}} 
\tikzset{triangle/.style={draw,regular polygon, regular polygon
    sides=3}} 
\tikzset{->-/.style={decoration={
  markings,
  mark=at position .5 with {\arrow{>}}},postaction={decorate}}} 
\tikzset{-<-/.style={decoration={ 
  markings,
  mark=at position .5 with {\arrow{<}}},postaction={decorate}}}
\newcommand\be{\begin{equation}}
\newcommand\ee{\end{equation}}
\newcommand\bea{\begin{eqnarray}}
\newcommand\eea{\end{eqnarray}}

\begin{document}

\begin{titlepage}
\renewcommand{\thefootnote}{\fnsymbol{footnote}}

\vspace*{1.0cm}

\begin{center}
{\textbf{\huge Rindler Observer Sublimation
}}
\end{center}
\vspace{1.0cm}

\centerline{
\textsc{\large J. A.  Rosabal}
\footnote{j.alejandro.rosabal@gmail.com}
}

\vspace{0.6cm}

\begin{center}
\it Asia Pacific Center for Theoretical Physics, Postech, Pohang 37673, Korea
\end{center}

\vspace*{1cm}

\centerline{\bf Abstract}

\begin{centerline}
\noindent

In this note, we propose that an object moving with proper constant acceleration, i.e., a Rindler observer experiences a sublimation (or evaporation) process. In this first proposal, we do not consider the backreaction due to the sublimation.
We focus on charged matter particles for the discussion, but for simplicity, we present the quantization of the neutrally charged massive scalar field in Rindler space. The amplitude from the Minkowski observer perspective of detection of matter particles that have been emitted by a Rindler observer, or accelerated detector, is computed in a new fashion.
We make a comparison between the Rindler observer sublimation and the black hole evaporation. We present three variants of a  new experimental setup, and we show that in two of them, the Minkowski amplitude of detection of matter particles corresponds to that of a thermal process. There is one, however, where deviations from thermality can be found. It is numerically explored.

\end{centerline}

\vspace*{1cm}

\noindent
 \textbf{Keywords:} Accelerated detectors, Unruh effect, Hawking radiation, QFT on accelerated frames.

\thispagestyle{empty}

\end{titlepage}

\setcounter{footnote}{0}

\tableofcontents

\newpage

\section{\label{sec:1}Introduction}

Black holes evaporation \cite{Hawking:1974rv, Hawking:1974sw, Hartle:1976tp, Hawking:1976ra} is perhaps the most fantastic prediction in modern physics. However, its experimental confirmation is far from being reached, mainly because of the probability of emission of a stellar black hole is ridiculously tiny. The lifetime, for instance, for a black hole of solar mass is much longer than the age of the universe \cite{Hawking:1974rv}, this is an indication that this process is possible but unlikely. Despite this unlikeliness, the efforts for understanding the phenomena related to the quantum mechanics of the black holes have been increasing over the last four decades. Nevertheless, so far, we do not have a full understanding of this phenomenon.

A related phenomenon that could offer some hints about the black hole quantum mechanics is the Unruh effect \cite{Fulling:1972md, Davies:1974th, Unruh:1976db}. A Rindler observer, an object with proper constant acceleration (we refer to it as accelerated detector too), experiences the vacuum as full of real pairs of entangled particle-antiparticle \cite{Rosabal:2018hkx, Rosabal:2018tbo}, and hence it perceives thermal radiation.

It turns out that under specific considerations, an object with proper constant acceleration might evaporate too. In this paper regard the Rindler observers as solid objects, and the final state of its constituents as a gas, for instance, of electrons, this is why we refer to this process as {\it Rindler observer sublimation}.

In this paper, we discuss the sublimation process of an accelerated detector without considering the backreaction due to the sublimation itself. We shall view the amplitude of the detection of particles emitted from an accelerated object, from the Minkowski perspective. Here we focus on electrons and positrons. We shall also discuss the interpretation of the amplitudes and probabilities in correspondence to the experiment. We will show that there are three possible experimental setups where we can get quantitatively different results related to the thermal nature of the radiation perceived by the Minkowski observer.

Our results are presented in parallel to reference \cite{Hartle:1976tp}. In the end, we conclude that the sublimation of a Rindler observer is not too different from the evaporation of a black hole. Although, no information paradox appears in the case under consideration.

As we show, and it is well known for electromagnetic radiation \cite{Unruh:1976db}, the probabilities of emission and absorption of a uniformly accelerated particle correspond to a thermal process. However, we will see how from the Minkowski observer point of view, there could be situations where deviations from thermality can be found. These deviations from thermality are intrinsically connected to the uniformly accelerated detectors and do not have any analog in black holes.

For motivating the idea we want to put forward, we would like to focus on the accelerated objects. First, let us point out that usually, the Unruh effect is related to electromagnetic fields. Nevertheless, it is universal and holds for any quantum field in nature. Either by field quantization or by using some accelerated detector model, the same effect comes out. Although, extra care is needed when considering detector models due to some discrepancies with the field quantization \cite{Sriramkumar:1999nw}.

Perhaps, the most popular detector models are the Unruh-DeWitt type \cite{Unruh:1976db, DeWitt}. When accelerated, they can absorb and emit electromagnetic radiation (usually modeled as massless scalar fields). If the detector made of matter particles, electrons, for instance, is found in a state different than its ground state at some time, it is said that the detector has detected a quanta of the electromagnetic field.

Now, if the Unruh effect is also valid for matter particles, we could use accelerated detectors for probing these kinds of background too. Note, however, that it is hard to apply the excited state argument to a matter particles detector after having absorbed a particle of the same kind. In other words, electrons, for instance, can absorb and emit photons but no electrons. So if we consider an accelerated detector made of electrons and the quantum vacuum that it is probing is a background of electrons (and positrons), absorption or emission yield to the gaining or losing of its constituents. In \cite{Fosco:2008vn, Fosco:2010vm} similar processes were described. We could also view this phenomenon as a tunneling process similar to \cite{Parikh:1999mf} for black holes.

Perhaps the simplest model we could build for a matter particle detector would be an accelerated box with a given number of particles (electrons, for instance) confined within it, with a very high confining potential. If, after some time, the number of particles within the box increases or decreases, it would be an indication that the detector has probed the vacuum. For this accelerated box, no interaction term between the detector and a quantum field \cite{Unruh:1976db, DeWitt} is needed.

Due to the similarity that the processes described above have with the sublimation process of a given substance, we have chosen to call it {\it sublimation} instead of evaporation. Notice that
when focusing only on the electromagnetic field, unlike for black holes, for uniformly accelerated detectors (made of matter particles), there is no room for introducing the idea of sublimation. \\

To derive the propagator between two points in the black hole geometry \cite{Hartle:1976tp}, Hartle and Hawking used the Feynman's worldline path integral (WPI) formulation \cite{Feynman:1950ir, Strassler:1992zr, Gies:1999jt}. They consider one point inside the horizon and the other outside. Two different patches are involved in the specification of these points location, also, in the specification of the initial and final state.

In this work, with the end of getting the amplitudes we are interested in, we work with the propagator between one point inside the Rindler wedge and the other point anywhere in Minkowski space. We use two different patches, too, as in \cite{Hartle:1976tp}. Here, we do not present the derivation of the Green's function using the WPI. Still, supported by it, we extract some useful information that allows us to make conclusions on which propagator is appropriate for obtaining the amplitudes.

\begin{figure}[h]
\centering
\includegraphics[width=.4\textwidth]{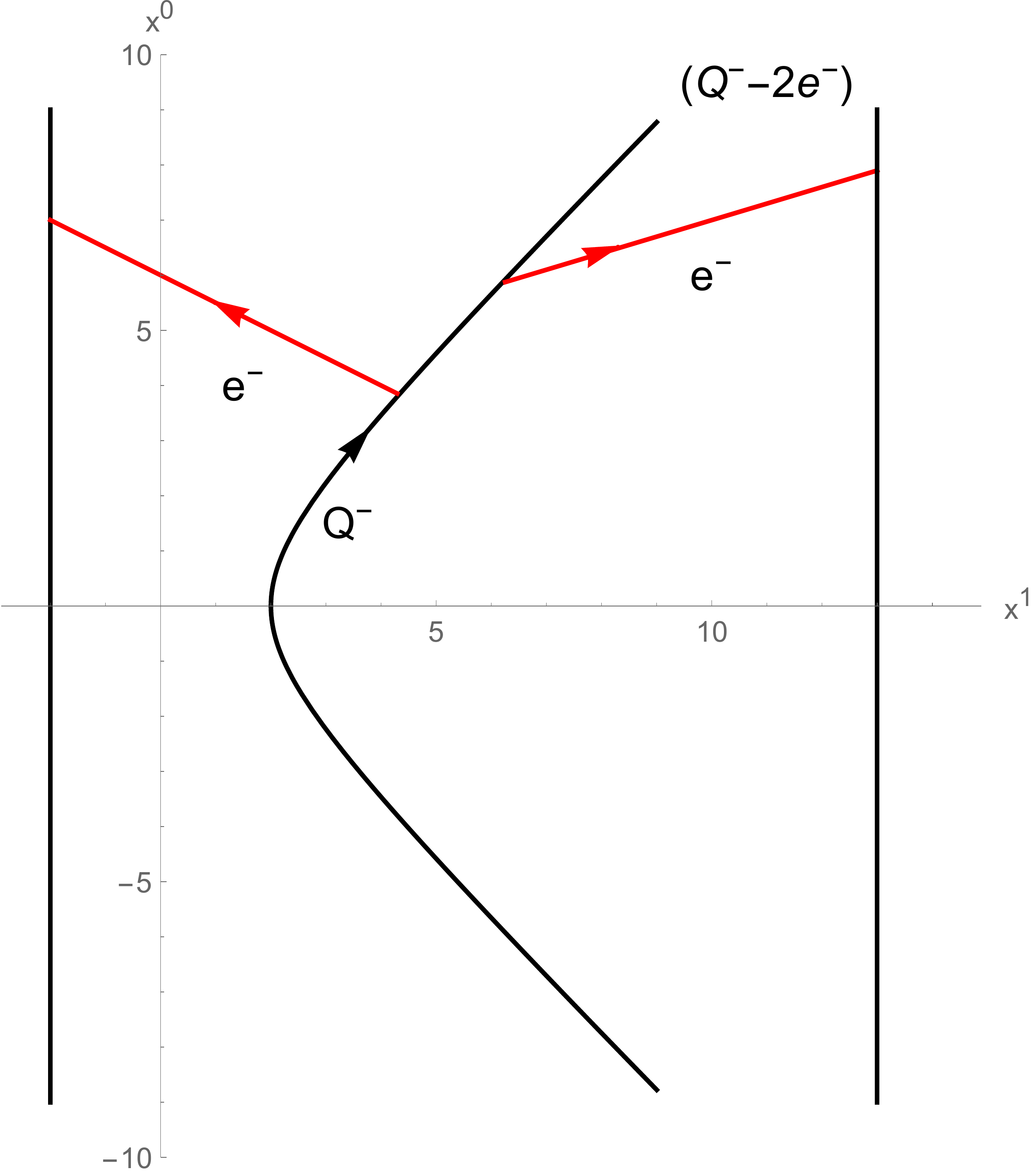}
\caption{\sl Pictorial representation of the Rindler observer sublimation. The red straight lines represent the stationary paths which connect the accelerated object and the observation points.}\label{detector_emission}
\end{figure}

The worldline path integral formulation offers an intuitive way of representing the emission and absorption processes. If one attempts to evaluate the amplitude by applying the method of stationary phase to the integrals involved, one finds that the stationary paths which connect the accelerated particle and the observation point are straight lines Fig.\ref{detector_emission}.

The processes depicted in Fig. \ref{detector_emission} are unlikely, like for black holes \cite{Hartle:1976tp}. We arrive at this conclusion after computing the amplitudes related to them. Although there is an experimental setup with a small window for confirmation. \\

The paper is organized as follows. In section \ref{sec:2}, the quantization in Rindler space using different boundary conditions to the ordinary ones in the canonical quantization is reviewed. The interpretation of the amplitudes is discussed in section \ref{sec:3}. In this section, a new derivation of the amplitudes is presented in parallel to reference \cite{Hartle:1976tp}. Three new experimental setups for detecting the radiation emitted from an accelerated detector are presented. In particular, one of them leads to a non-thermal emission. Strictly speaking, it could be called ``a non-thermal detection by the Minkowski observer,'' this is clarified in \ref{subsec:3.1}. The non-thermal radiance is numerically explored in \ref{subsec:3.2}. After conclusions, two appendixes are presented.

 \section*{Notation and conventions}

We use $\varphi_R$ to denote the scalar field in Rindler coordinates, while $\varphi_M$ denotes the scalar field in Minkowski coordinates. $x^0 (\tau,\rho) = \rho\ \text{sinh}(\tau)$, and $x^1(\tau,\rho) = \rho\ \text{cosh}(\tau)$, are used to denote the relation between the Minkowski and Rindler coordinates. The normalized eigenfunctions in the $\rho$ direction in Rindler space are $\psi_{\nu,\kappa}(\rho)$. Where $\nu$ is a quantum number associated to the energy in the Rindler frame; and $\kappa=(|\bar{k}|^2+m^2)^{\frac{1}{2}}$, where $\bar{k}$ is de momentum in the $(x_1,x_2)$ directions, being $m$, the mass of the scalar particles. We refer to the amplitudes and probabilities of emission and absorption of the accelerated detector as $A_{emi}$, and $A_{abs}$; $P_{emi}$, and $P_{abs}$, respectively. At the same time, they can be regarded as the amplitudes and probabilities of absorption and emission of a screen placed at some distance relative to the accelerated detector. We refer to this screen as a Minkowski detector. $J_{\mu}(x)$, denotes the quantum mechanics current of probability. While $\varphi_1$, and $\varphi_2$, are the quantum mechanics wave functions as perceived from the Rindler and the Miskowski observer perspective.

\section{\label{sec:2} Quantization}

In this section, we review the quantization of a massive scalar field in Rindler space following \cite{Rosabal:2018hkx, Rosabal:2018tbo}. For simplicity, we work with a massive neutrally charged scalar field, but this work can be easily extended to massive charged scalars or fermions.

In \cite{Rosabal:2018hkx, Rosabal:2018tbo} in contrast to the ordinary canonical quantization, the field operator is subjected to the boundary conditions
\bea\label{B.C}
\varphi_R(\tau_i,\rho,\bar{x}) & = & \varphi_M(x^0(\tau_i,\rho),x^1(\tau_i,\rho),\bar{x}), \nonumber\\
\varphi_R(\tau_f,\rho,\bar{x}) & = & \varphi_M(x^0(\tau_f,\rho),x^1(\tau_f,\rho),\bar{x}),
\eea
where
\bea\label{transf0}
x^0 (\tau,\rho) & = & \rho\ \text{sinh}(\tau), \nonumber\\
x^1(\tau,\rho) & = & \rho\ \text{cosh}(\tau).
\eea
These are the boundary conditions that are consistent with the Lagrangian approach of field theory. In the Rindler frame, $\varphi_R$, is specified at the initial and final time $\tau_i, \tau_f$, respectively. These are the slices that intersect the points where the acceleration is turned on and off. The operator $\varphi_M$, can be obtained in the canonical quantization in Minkowski space from the solution of
\be
(\Box-m^2) \varphi_M(x^0,x^1,\bar{x}) = 0. \label{Mink.eom}
\ee

The massive scalar field action in the Rindler wedge $\tau_i<\tau<\tau_f$ reads
\begin{multline}
S=-\frac{1}{2}\int\limits_{\tau_i}^{\tau_f} d\tau \int\limits_{0}^{\infty}d\rho \int\limits_{-\infty}^{\infty} d^2x\Big[-\rho^{-1}(\partial_{\tau}\varphi_{R})^2\\
+\rho\big((\partial_{\rho}\varphi_{R})^2+(\partial_{x^2}\varphi_{R})^2+(\partial_{x^3}\varphi_{R})^2+ m^2 \varphi_{R}^2 \big) \Big]\label{RindlerAction}.
\end{multline}

The variation of \eqref{RindlerAction} subjected to $\delta\varphi_{R}(\tau_i,\rho,\bar{x})=\delta\varphi_{R}(\tau_f,\rho,\bar{x})=0$, and $\delta S=0$, leads to
\be
(\Box-m^2) \varphi_R(\tau,\rho,\bar{x}) = 0\label{Rindler.eom}.
\ee

The general solution of \eqref{Mink.eom} is
\begin{multline}\label{modeexM}
\varphi_M(x^0,x^1,\bar{x}) =
\int\limits_{-\infty}^{\infty}\frac{dk_1}{2\pi}\int\limits_{-\infty}^{\infty}\frac{d^2k}{(2\pi)^2}\frac{1}{(2k_0)^{\frac{1}{2}}}\\
\Big(a_{(k_1,\bar{k})}\text{e}^{-\text{i}(k_0x^0+k_1x^1+\bar{k}\cdot\bar{x})}
+a^{\dagger}_{(k_1,\bar{k})}\text{e}^{\text{i}(k_0x^0+k_1x^1+\bar{k}\cdot\bar{x})}\Big).
\end{multline}
While the general solution of \eqref{Rindler.eom} reads \cite{Fulling:1972md}
\be\label{modeexR}
\varphi_R(\tau,\rho,\bar{x})=\int\limits_{0}^{\infty}d\nu\int\limits_{-\infty}^{\infty}\frac{d^2k}{(2\pi)^2}\frac{1}{(2\nu)^{\frac{1}{2}}}
\Big(b_{(\nu,-\bar{k})}\text{e}^{-\text{i}\nu \tau}+b^{\dagger}_{(\nu,\bar{k})}\text{e}^{\text{i}\nu \tau}\Big)\psi_{\nu,\kappa}(\rho)\text{e}^{\text{i}\bar{k}\cdot\bar{x}},
\ee
with
\be
\kappa=(|\bar{k}|^2+m^2)^{\frac{1}{2}},\label{kappa}
\ee
and
\be
k_0=(k_1^2+|\bar{k}|^2+m^2)^{\frac{1}{2}}=(k_1^2+\kappa^2)^{\frac{1}{2}},
\ee
$|\bar{k}|^2=k_2^2+k_3^2$, and $\psi_{\nu,\kappa}(\rho)$, the normalized eigenfunctions of the equation
\be
\Big(\rho^2\frac{d^2}{d\rho^2}+\rho\frac{d}{d\rho}-\kappa^2 \rho^2+\nu^2\Big)\psi_{\nu,\kappa}(\rho)=0,\label{function}
\ee
\be
\psi_{\nu,\kappa}(\rho)=\pi^{-1}\big(2\nu \text{sinh}(\pi \nu)\big)^{\frac{1}{2}}\text{K}_{\text{i}\nu}(\kappa\rho),
\ee
\be
\int\limits_0^{\infty}\frac{d\rho}{\rho}\ \psi_{\nu,\kappa}(\rho)\psi_{\nu^{\prime},\kappa}(\rho)=\delta(\nu-\nu^{\prime}),
\ee
where $\text{K}_{\text{i}\nu}(\kappa\rho)$, are the modified Bessel function of the second kind.

Imposing \eqref{B.C} we get the equations
\be
\text{e}^{-\text{i}\nu \tau_{i}}b_{(\nu,-\bar{k})}+\text{e}^{\text{i}\nu \tau_{i}}b^{\dagger}_{(\nu,\bar{k})}=
(2\nu)^{\frac{1}{2}}\int\limits_{0}^{\infty}d\rho\int\limits_{-\infty}^{\infty}d^2x\frac{\psi_{\nu,\kappa}(\rho)}{\rho}\text{e}^{-\text{i}\bar{k}\cdot\bar{x}}\varphi_{M}(\tau_{i},\rho,\bar{x}),
\ee
and
\be
\text{e}^{-\text{i}\nu \tau_{f}}b_{(\nu,-\bar{k})}+\text{e}^{\text{i}\nu \tau_{f}}b^{\dagger}_{(\nu,\bar{k})}=
(2\nu)^{\frac{1}{2}}\int\limits_{0}^{\infty}d\rho\int\limits_{-\infty}^{\infty}d^2x\frac{\psi_{\nu,\kappa}(\rho)}{\rho}\text{e}^{-\text{i}\bar{k}\cdot\bar{x}}\varphi_{M}(\tau_{f},\rho,\bar{x}),
\ee
we have used the short hand notation
\be
\varphi_M(\tau,\rho,\bar{x}) = \varphi_M(x^0(\tau,\rho),x^1(\tau,\rho),\bar{x}).
\ee
Solving for $b_{(\nu,\bar{k})}$, we get
\begin{multline}\label{coefficient.b}
b_{(\nu,\bar{k})} = -\text{i}\frac{(2\nu)^{\frac{1}{2}}}{2\text{sin}\big[\nu(\tau_f-\tau_i)\big]}\int\limits_{0}^{\infty}d\rho\int\limits_{-\infty}^{\infty}d^2x\\
\frac{\psi_{\nu,\kappa}(\rho)}{\rho}\text{e}^{\text{i}\bar{k}\cdot\bar{x}}\Big[
\text{e}^{\text{i}\nu\tau_f}\varphi_{M}(\tau_i\ ,\rho,\bar{x}) -\text{e}^{\text{i}\nu\tau_i} \varphi_{M}(\tau_f\ ,\rho,\bar{x}) \Big].
\end{multline}

In \cite{Rosabal:2018hkx, Rosabal:2018tbo}, an alternative way of deriving the vacuum energy was presented without using the Bogoliubov coefficients. It highlights the loops and the open paths contributions to the Rindler vacuum energy.

Here we shall present the Bogoliubov coefficients derived from the boundary conditions \eqref{B.C}. It will be surprising for the reader, despite we are using different boundary conditions to \eqref{B.C.c.q}, from \eqref{B.C} we get exactly the Bogoliubov coefficients we can obtain from the canonical quantization \cite{Fulling:1972md}, see Appendix \ref{appen:A}.

For this purpose we solve integrals of the form
\be
\int\limits_{0}^{\infty}d\rho\int\limits_{-\infty}^{\infty}d^2x\\
\frac{\psi_{\nu,\kappa}(\rho)}{\rho}\text{e}^{\text{i}\bar{k}\cdot\bar{x}}
\varphi_{M}(\tau,\rho,\bar{x}),
\ee
which equals to
\begin{multline} \label{Int.coeff}
\frac{1}{4\pi\nu^{\frac{1}{2}}}\frac{1}{(\text{sinh}(\pi\nu))^{\frac{1}{2}}}\int\limits_{-\infty}^{\infty}d\tilde{z}\Big (\\
a_{(z,\bar{k})}\big[(\text{i})^{-\text{i}\nu}\text{e}^{-\text{i}\nu(\tau+z)}+(\text{i})^{\text{i}\nu}\text{e}^{\text{i}\nu(\tau+z)} \big]\\
+a^{\dagger}_{(z,-\bar{k})}\big[(\text{i})^{-\text{i}\nu}\text{e}^{\text{i}\nu(\tau+z)}+(\text{i})^{\text{i}\nu}\text{e}^{-\text{i}\nu(\tau+z)} \big]\Big),
\end{multline}
where
\be
z=\text{arcsinh}(\frac{p1}{\kappa}),\label{rela1}
\ee
and
\be
d\tilde{z}=\kappa^{\frac{1}{2}}(\text{cosh}(z))^{\frac{1}{2}} dz=\frac{dp_1}{(\sqrt{p_1^2+\kappa^2})^{\frac{1}{2}}}.\label{rela2}
\ee
Plugging \eqref{Int.coeff} in \eqref{coefficient.b} we get
\be\label{pre-Bogo}
b_{(\nu,\bar{k})}=\frac{1}{2\pi}\frac{1}{(\text{e}^{2\pi\nu}-1)^{\frac{1}{2}}}\int\limits_{-\infty}^{\infty}d\tilde{z}\Big(\text{e}^{\pi\nu} a_{(z,\bar{k})}+ a^{\dagger}_{(z,-\bar{k})}\Big)\text{e}^{-\text{i}\nu z}.
\ee
Using \eqref{rela1} and \eqref{rela2} we arrive at \eqref{Bogoliubov coefficients} which are the Bogoliubov coefficients as presented, for instance, in \cite{Fulling:1972md}. We emphasize that here they have been obtained from \eqref{B.C}.

The vacuum energy $E_{vac}^R$ in Rindler space is given by
\be
E_{vac}^R=\int\limits_{-\infty}^{\infty} \frac{d^2k}{(2\pi)^2}\int\limits_{0}^{\infty} d\nu \nu \langle 0^M | b^{\dagger}_{(\nu,\bar{k})} b_{(\nu,\bar{k})}|0^M\rangle\ +E^{R}_{0},\label{Rindler.V.E}
\ee
where
\be
E^{R}_{0} = \frac{1}{2}\delta(0)^3 \int\limits_{-\infty}^{\infty}d^2k \int\limits_{0}^{\infty}d\nu \nu,
\ee
is the contribution of the loops inside the Rindler wedge, as discussed in \cite{Rosabal:2018hkx}. Usually, this contribution is discarded. The Minkowski vacuum $|0^M\rangle$, satisfies $a_{(k_1,\bar{k})}|0^M\rangle=0$. The thermal distribution can be easily computed using \eqref{Bogoliubov coefficients} and \eqref{deltafunct},
\be
E^{R}_{vac}= \int\limits_{-\infty}^{\infty}\frac{d^2k}{(2\pi)^2} \int\limits_{0}^{\infty}\frac{d\nu}{2\pi} \frac{\nu}{\text{e}^{2\pi \nu}-1}V_3+E_0^R,
\ee
where
\be
2\pi\delta(0)=\int\limits_{-\infty}^{\infty}dx=V_1, \ V_1^3=V_3.\label{delta_volumen}
\ee

\section{\label{sec:3} Rindler Observer Sublimation}

The accelerated detector at some initial time can be idealized as a solid made of electrons (and other particles). After some time, all these electrons would be part of the radiation. Namely, they could be considered as a gas of electrons. How likely could this process be? We answer this question in the next section by computing the probability of emission and absorption of a uniformly accelerated detector from the Minkowski perspective.

\subsection{\label{subsec:3.1} Amplitudes and Thermal Radiance}

Let us now compute the amplitude associated with the emission or absorption of one Rindler mode by the accelerated detector from the Minkowski observer perspective. Namely, the amplitude for the process in Fig. \ref{detector_emission}. This particular amplitude tells us, like in \cite{Hartle:1976tp}, for a black hole, whether or not a Rindler observer can emit matter particles.

It is useful to know that if the accelerated detector emits a Rindler mode, we regard the initial state as $b^{\dagger}_{(\nu,\bar{k})}|0^M\rangle$. Conversely, if a Rindler mode is absorbed by the accelerated detector the initial state is $ b_{(\nu,\bar{k})}|0^M\rangle$.

The amplitudes of the process described above, where the final state perceived by a Minkowski observer is a Minkowski particle $a^{\dagger}_{(k_1,\bar{k})}|0^M\rangle$, are
\be\label{ampli1}
A_{ b^{\dagger}\to a^{\dagger}}= \langle 0^M | a_{(k^{\prime}_1,\bar{k}^{\prime})} b^{\dagger}_{(\nu,\bar{k})}|0^M\rangle=\frac{\text{e}^{\pi\nu}}{(\text{e}^{2\pi\nu}-1)^{\frac{1}{2}}}\frac{1}{\sqrt{k^{\prime}_0}}
(\frac{k^{\prime}_0+k^{\prime}_1}{k^{\prime}_0-k^{\prime}_1})^{\frac{1}{2}\text{i}\nu}(2\pi)^2\delta^2({\bar{k}^{\prime}-\bar{k}}),
\ee
and
\be \label{ampli2}
A_{ b\to a^{\dagger}}= \langle 0^M | a_{(k^{\prime}_1,\bar{k}^{\prime})} b_{(\nu,\bar{k})}|0^M\rangle=\frac{1}{(\text{e}^{2\pi\nu}-1)^{\frac{1}{2}}}\frac{1}{\sqrt{k^{\prime}_0}}
(\frac{k^{\prime}_0+k^{\prime}_1}{k^{\prime}_0-k^{\prime}_1})^{-\frac{1}{2}\text{i}\nu}(2\pi)^2\delta^2({\bar{k}^{\prime}-\bar{k}}),
\ee
where $ k_0=(k_1^{\prime 2}+|\bar{k}^{\prime }|^{2}+m^2)$. Here we have used \eqref{Bogoliubov coefficients}.

The total probability will be the sum over all initial modes for a given frequency $\nu$, \cite{Hartle:1976tp} of the square of the amplitude. A Rindler mode $b^{\dagger}_{(\nu,\bar{k})}|0^M\rangle$ or $b_{(\nu,\bar{k})}|0^M\rangle$ is fully specified by three quantum number $(\nu, k_2,k_3)$. For a definite frequency mode the total amplitude is
\bea\nonumber\label{Total_Proba}
P_{emi} & = &\int\limits_{-\infty}^{\infty}\frac{d^2\bar{k}}{(2\pi)^2}\frac{|A_{emi}|^2}{V_2},\\
P_{abs} & = & \int\limits_{-\infty}^{\infty}\frac{d^2\bar{k}}{(2\pi)^2}\frac{|A_{abs}|^2}{V_2}.
\eea
The formal square of the $\delta$ function is considered as
\be
\delta^2({\bar{k}^{\prime}-\bar{k}})=\frac{V_2}{(2\pi)^2}\delta({\bar{k}^{\prime}-\bar{k}}),
\ee
where we have used \eqref{delta_volumen}. One can avoid these formal manipulations using wave packets, see for instance \cite{JJ}.

Notice that the total probability fulfill the relation \cite{Hartle:1976tp, Padmanabhan:2019yyg}
\be
\frac{P_{abs}}{P_{emi}}=\text{e}^{-2\pi\nu},\label{thermal_ratio}
\ee
which indicates that the emission and absorption process is thermal with a temperature
\be
T=\frac{\hbar \bold{a}}{2\pi\text{c} k_B}, \label{temp}
\ee
we have restored the acceleration $\bold{a}$ and the physical constants. Relation \eqref{thermal_ratio} implies that the probability of absorption is always smaller than the probability of emission. Namely, for the accelerated detector, it is more likely emit than absorb. Hence, gradually it loses its constituents. We recall at this point that we are considering emission and absorption of matter particles, although the same analysis goes to electromagnetic radiation.

We want to make a parenthesis here to discuss what transition amplitudes we are computing in \eqref{ampli1} and \eqref{ampli2}. The final Minkowski state involved in \eqref{ampli1} and \eqref{ampli2} is defined over a space-like slice at the Minkowski time $x^0=+\infty$. So, it means that if we wait long enough, there will be transitions from Rindler modes emitted by the accelerated object to Minkowski particles. However, \eqref{ampli1} and \eqref{ampli2} do not tell us how and where to measure in the three-dimensional space to detect these particles.

In what follows, we present a different derivation of the radiance of a Rindler observer. It makes transparent how and where we should measure in the three-dimensional space to detect some radiated particles. This calculation is made in analogy to the one presented in \cite{Hartle:1976tp} for the black hole radiance. It shows that the radiative processes of a Rindler observer are no so different from those in a black hole. In the end, we conclude that in the same way a black hole evaporates\footnote{We refer the reader to section IV of \cite{Hartle:1976tp}.} \cite{Hartle:1976tp} a Rindler observer sublimates. \\

As we are dealing with the emission and absorption of one single particle, we can use the probability current
\be
\text{J}_{\mu}(x)=-\text{i}\Big(\varphi^{*}_2(x)\partial_{x^\mu}\varphi(x)-\partial_{x^\mu}\varphi^{*}_2(x)\varphi(x) \Big),\label{flux}
\ee
to compute the amplitude. By proceeding in this way, we will gain some intuition on how to detect the radiated particles.

\begin{figure}[h]
\centering
\includegraphics[width=.5\textwidth]{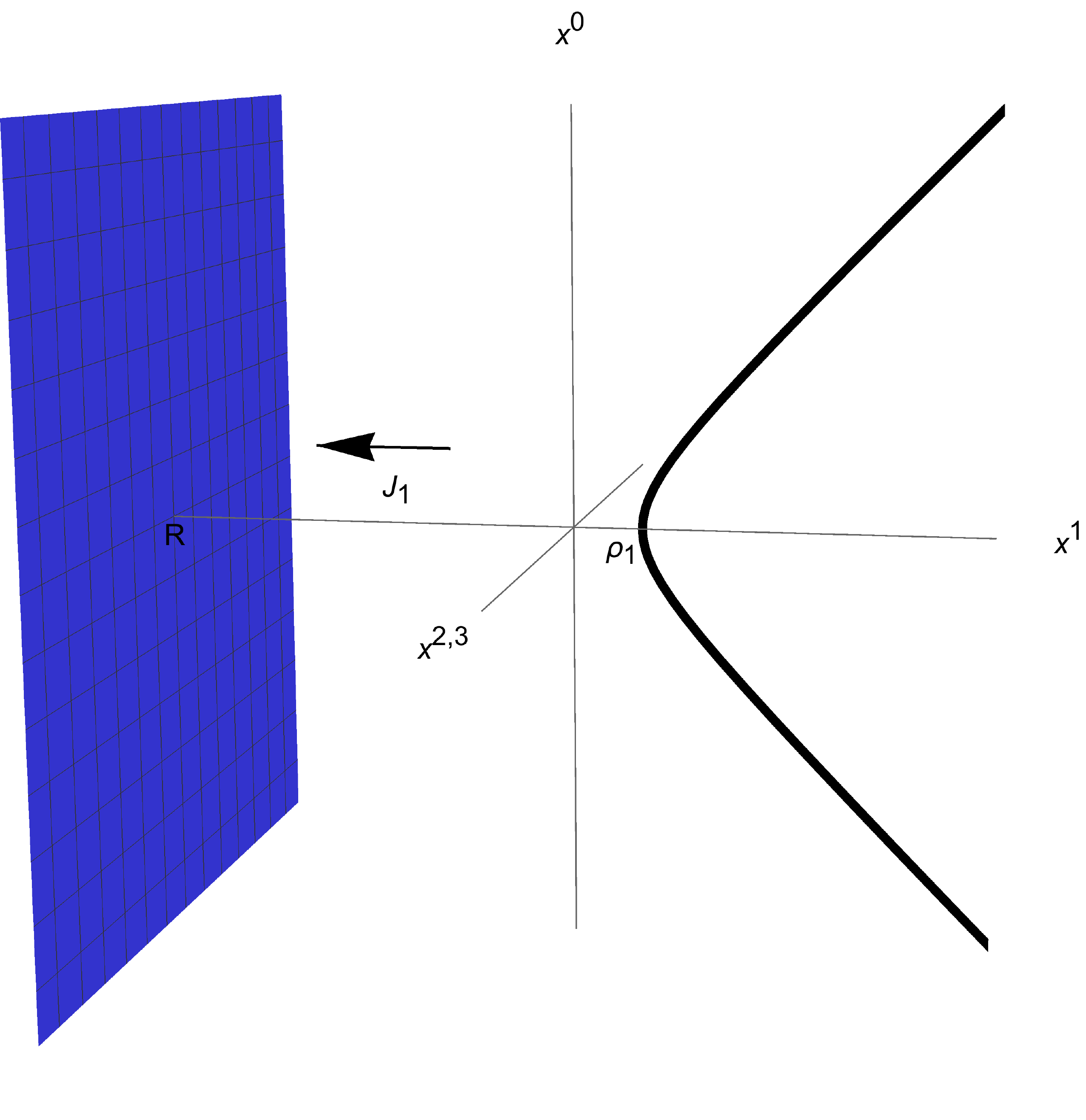}
\caption{\sl Pictorial representation in spacetime of the experimental setup. In blue, the screen (Minkowski detector). The thick black line represents the accelerated object. In this setup, the screen is placed to the left of $x^1=\rho_1$. $\text{J}_{1}$, represents the current associated with an incoming wave moving from right to left.}\label{fig1}
\end{figure}

\begin{figure}[h]
\centering
\includegraphics[width=.5\textwidth]{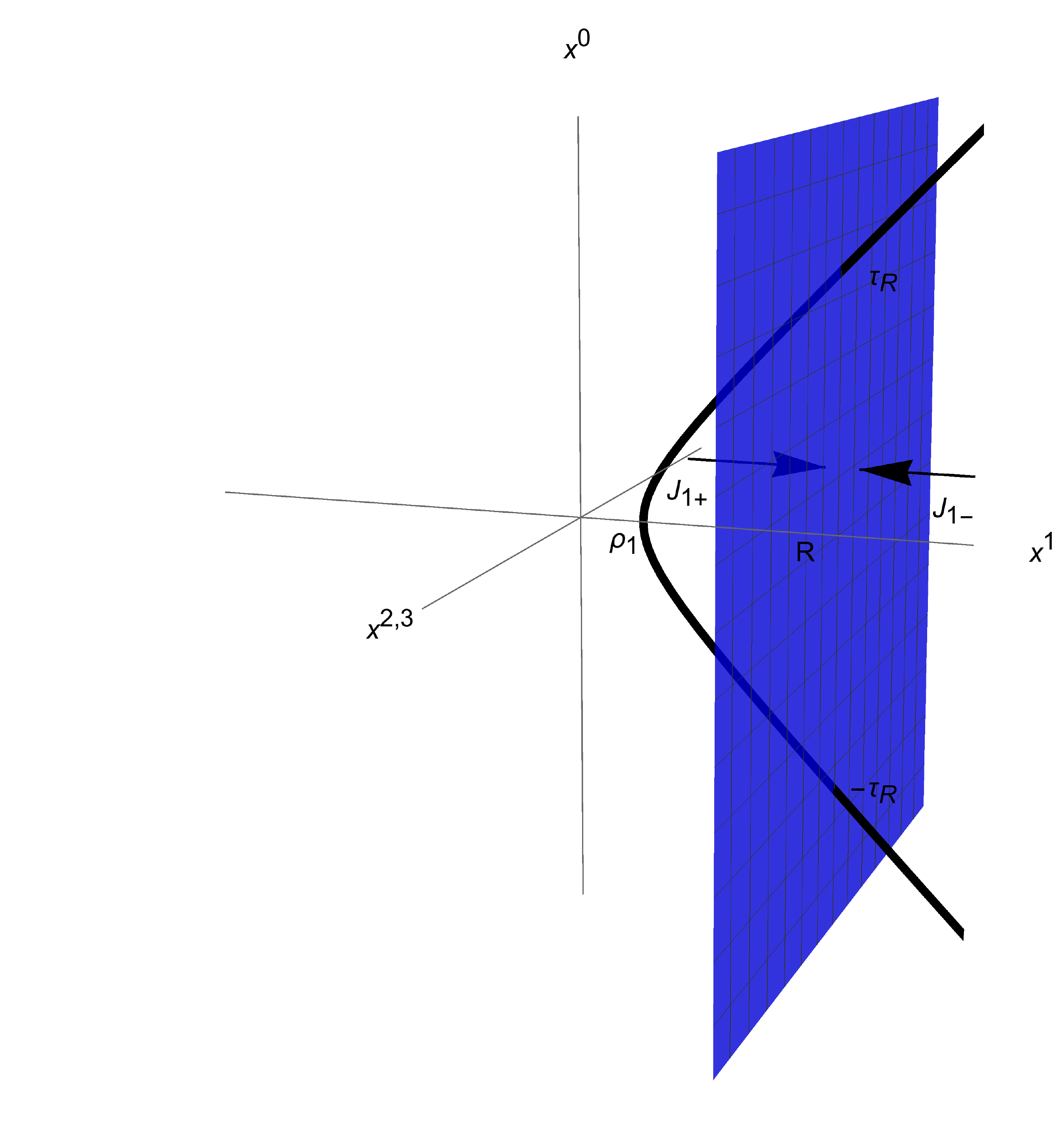}
\caption{\sl Pictorial representation in spacetime of the experimental setup. In blue screen (Minkowski detector). The thick black line represents the accelerated object. In this setup, the screen is placed to the right of $x^1=\rho_1$. $\text{J}_{1-}$, and $\text{J}_{1+}$, represent the current associated with the incoming waves on each side of the screen. $\tau_{R}$, is the Rindler time where the accelerated particle meets the screen}\label{fig2}
\end{figure}

First, suppose we place a screen \footnote{The screen is the analogous of the spherical detector placed outside the horizon to collect the Hawking radiation at a constant $r$ \cite{Hartle:1976tp}.} which is our radiation detector (Minkowski observer or detector) perpendicular to the direction of motion of the accelerated object, as indicated in Fig. \ref{fig1} and Fig. \ref{fig2}.

This screen measures on shell $k_0^2=k_1^2+|\bar{k}|^2+m^2$, purely positive energy particles in modes $\varphi_2(x)$. The location of the screen will be specified through the calculation. We will consider three situations. The screen placed to the left/right of $x^1=\rho_1$, and $R\rightarrow+\infty$.

The amplitude of detecting a mode $\varphi_2(x)$, at the screen $x^1=R$, from the Minkowski perspective, having started at $\rho=\rho_1$, in a mode $\varphi_1(y)$, from the Rindler perspective is given by the total flux of probability through the screen
\be\label{H-ampli}
A=\int\limits_{\text{screen}}\text{J}_{\mu}(x)\text{n}^{\mu}=\int\limits_{-\infty}^{\infty}dx^0\int\limits_{-\infty}^{\infty}d^2\bar{x}\ \text{J}_1(x)\text{n}^1
\Big|_{x^1=R},
\ee
where, using the Green's third identity,
\be\label{H-sol}
\varphi(x)=\rho_1\int\limits_{-\infty}^{\infty}d\tau\int\limits_{-\infty}^{\infty}d^2\bar{y}
\Big(\varphi_1(y)\partial_{\rho}G(x,y)-G(x,y)\partial_{\rho}\varphi_1(y) \Big)\Big|_{\rho=\rho_1},
\ee
and $\text{n}^{\mu}=(0,1,0,0)$, is the normal vector to the screen.
Here the coordinates $x$ are referred to the Minkowski observer, $x=(x^0,x^1,\bar{x})$, while $y=(\tau, \rho,\bar{y})$, are the Rindler coordinates. The appearance of $\rho_1$ in front of the integral \eqref{H-sol} is due to the volume measure on the surface $\rho=\rho_1$, i.e., $ds^2=-\rho^2d\tau^2+d\rho^2+d\bar{y}^2$, hence $\sqrt{-h}\Large|_{\rho=\rho_1}d\tau d^2\bar{y}=\rho_1d\tau d^2\bar{y}$. The surface $\rho=\rho_1$, could be taken anywhere inside the Rindler wedge, in a similar fashion to \cite{Hartle:1976tp} where the surrounding surface inside the horizon in taken at a constant $r$, for simplicity we place it on the same location of the accelerated particle. The Green's function $G(x,y)$, has one of its legs evaluated only in the Rindler wedge.

Since one point is in the right Rindler wedge and the other could be outside the wedge, the question now is, what Green's function should we use in \eqref{H-sol}. Note that if we were to compute a process involving only points inside the right Rindler wedge, we could use the thermal Green's function in Rindler space.

To answer the previous question, we can proceed as in \cite{Hartle:1976tp}. In this reference, one of the points where the Green's function is evaluated resides inside the horizon while the other could be anywhere in the Schwarzschild space\footnote{ Although, in \cite{Hartle:1976tp} the relevant portion of the space where the second point of the Green's function is evaluated is the exterior part of the black hole.}.

For deriving the Green's function between two points (no matter the location of these points) one can use its worldline path integral representation and the method of stationary phase \cite{Hartle:1976tp, Gies:1999jt}
\bea
G(x,y) & = & \int\limits_{0}^{\infty}ds\ \text{e}^{-\text{i}m^2s}
\int\limits_{x(0)=y}^{x(1)=x}Dx^{\mu}(\tau)\text{exp}\Big[\text{i}\frac{1}{4s}\int\limits_{0}^{1}d\tau\ \dot{x}^2(\tau)\Big]\nonumber\\
{} & = &\int_{0}^{\infty}ds\ \text{e}^{-\text{i}m^2s} \frac{1}{(2\pi \text{i})^2}\sqrt{D(x,y,s)}\text{e}^{\text{i}S(x,y,s)}, \label{wpi_repre}
\eea
where $S(x,y,s)$ is the action evaluated on the classical path connecting $y$ and $x$, and
\be
D(x,y,s)=\text{det}\Big( \frac{\partial}{\partial_{x^{\mu}}} \frac{\partial}{\partial_{y^{\nu}}} S(x,y,s) \Big)
\ee
The point $x$ belongs to the Minkowski patch. The point $y$ belongs to the Rindler patch, but this patch is just a part of Minkowski space. So, we can regard $y$ as belonging to Minkowski space. With this in mind, the path integration in \eqref{wpi_repre} should be over all the path starting at $y$ and ending at $x$ in whole Minkowski space. The derivation of $G(x,y)$ from \eqref{wpi_repre} is a well known calculation, see for instance \cite{Feynman:1950ir, Strassler:1992zr, Gies:1999jt}. For this case the method of stationary phase simply reduces to the evaluation of the action on the straight line connecting the emission and observation points Fig. \ref{detector_emission}, it leads to
\be
G(x,y)=\int\limits_{-\infty}^{\infty}\frac{d^4p}{(2\pi)^4}\frac{\text{e}^{-\text{i}\big((x^0-y^0)p_0+(x^1-y^1)p_1+(\bar{x}-\bar{y})\cdot\bar{p}\big)}}{p^2-m^2+\text{i}\epsilon},
\ee
in our case, with $y$ restricted to the right wedge, i.e.,
\bea\label{transf}
y^0 & = & \rho\ \text{sinh}(\tau),\nonumber \\
y^1 & = & \rho\ \text{cosh}(\tau).
\eea

Relations \eqref{flux}, \eqref{H-ampli} and \eqref{H-sol} can be combined in a more compact form
\be
A=-\int d\sigma^{\mu}(x)\int d\sigma^{\nu}(y)\varphi_2^{*}(x)\overset\leftrightarrow{\partial}_{\mu}G(x,y)\overset\leftrightarrow{\partial}_{\nu}\varphi_1(y),\label{compactA}
\ee
where the integral over $x$ is taken over the surface $x^1=R$, and the integral over $y$ is over the surface $\rho=\rho_1$, and $a\overset\leftrightarrow{\partial}_{\mu}b=a\partial_{\mu}b-b\partial_{\mu}a$. This formula is similar to $(4.1)$ of reference \cite{Hartle:1976tp}.

In order to further proceed we need to find $\varphi_1(y)$, and $\varphi_2(x)$. They are the quantum mechanical wave functions. To be consistent with our conventions, we use the mode expansion \eqref{modeexR} together with \eqref{Bogoliubov coefficients} to compute the wave functions. The emission wave function from the Rindler point of view is given by
\bea\nonumber\label{emiWF}
\varphi_1(y) & = &\langle 0^M|\varphi_R(\tau,\rho,\bar{y}) b^{\dagger}_{(\nu,\bar{k})}|0^M\rangle\\
{} & = & \text{N}_1\ \text{e}^{-\text{i}\nu\tau}\text{K}_{\text{i}\nu}(\kappa\rho)\text{e}^{-\text{i}\bar{k}\cdot\bar{y}},
\eea
where $\text{N}_1=\frac{1}{\sqrt{2}\pi}\frac{\text{e}^{\frac{3}{2}\pi\nu}}{(\text{e}^{2\pi\nu}-1)^{\frac{1}{2}}}$.
On the other hand the absorption wave function is
\bea\nonumber\label{absWF}
\varphi_1(y) & = &\langle 0^M|\varphi_R(\tau,\rho,\bar{y}) b_{(\nu,\bar{k})}|0^M\rangle\\
{} & = & \text{N}_1^{\prime} \ \text{e}^{\text{i}\nu\tau}\text{K}_{\text{i}\nu}(\kappa\rho)\text{e}^{\text{i}\bar{k}\cdot\bar{y}},
\eea
where $\text{N}_1^{\prime}=\frac{1}{\sqrt{2}\pi}\frac{\text{e}^{-\frac{1}{2}\pi\nu}}{(\text{e}^{2\pi\nu}-1)^{\frac{1}{2}}}$. Notice that they have support only on the right Rindler wedge.
Finally, the Minkowski wave function is simple given by
\bea\nonumber\label{MWF}
\varphi_2(x) & = &\langle 0^M|\varphi_M(x) a^{\dagger}_{(k_1,\bar{k})}|0^M\rangle\\
{} & = & \frac{1}{\sqrt{2k_0}}\text{e}^{-\text{i}kx}=\text{N}_2\ \text{e}^{-\text{i}kx}.
\eea
We have used the relation
\be
\int\limits_{-\infty}^{\infty}\frac{dp_1}{p_0}\Big(\frac{p_0+p_1}{p_0-p_1}\Big)^{\frac{1}{2}\text{i}(\nu^{\prime}-\nu)}=(2\pi)\delta(\nu^{\prime}-\nu)\label{deltafunct},
\ee
which can be easily proved by the change of variable $z=\frac{1}{2}\text{Log}\Big(\frac{p_0+p_1}{p_0-p_1}\Big)$. Although the Rindler wave functions have support only on the right wedge they can be written as localized (inside Rindler space) Minkowski wave package \eqref{wfRM1} and \eqref{wfRM2}.

Let us present the calculation of the emission amplitude. For this we use \eqref{emiWF} and \eqref{MWF}. After some algebra from \eqref{compactA} one arrives at
\begin{multline}\label{Amp-inter-res1}
A_{emi}=\frac{1}{2} \text{N}_1\text{N}_2\rho_1\text{e}^{\text{i}k^{(2)}_1R}(2\pi)^2\delta^2(\bar{k}^{(2)}-\bar{k}^{(1)})
\Big(\text{K}_{\text{i}\nu}(\kappa \rho_1)\partial_{\rho}- \partial_{\rho}\text{K}_{\text{i}\nu}(\kappa \rho_1) \Big)\Big(\partial_{x^1}-\text{i}k_1^{(2)} \Big)\\
\int\limits_{-\infty}^{\infty}d\tau \frac{\text{e}^{-\text{i}|x^1-y^1||k_1^{(2)}|}}{\big|k_1^{(2)}\big|}\text{e}^{\text{i}k_0^{(2)}y^0-\text{i}\nu\tau}\Big|_{\begin{matrix} \rho=\rho_1\\ x^1=\small{R}\end{matrix}},
\end{multline}
recall that $y^0$ and $y^1$ are given in \eqref{transf}.

We shall consider first the case depicted in Fig. \ref{fig1}, i.e., $x^1=R<\rho_1$, which implies that $x^1-y^1=R-\rho_1\ \text{cosh}(\tau)<0$, for all $\tau$. Also, an incoming wave moving from right to left with $k_1^{(2)}<0$, which represents a particle emitted by the accelerated object. Under these considerations \eqref{Amp-inter-res1} reduces to
\begin{multline}\label{Amp-inter-res2}
A_{emi}= \text{i}\text{N}_1\text{N}_2\rho_1(2\pi)^2\delta^2(\bar{k}^{(2)}-\bar{k}^{(1)})
\Big(\text{K}_{\text{i}\nu}(\kappa \rho_1)\partial_{\rho}- \partial_{\rho}\text{K}_{\text{i}\nu}(\kappa \rho_1) \Big)\\
\int\limits_{-\infty}^{\infty}d\tau \text{e}^{\text{i}k_1^{(2)}y^1+\text{i}k_0^{(2)}y^0-\text{i}\nu\tau}\Big|_{ \rho=\rho_1},
\end{multline}
where no $R$ dependence appears.
Being careful we can analytically extend the previous integral, see Appendix \ref{appen:B}, to get
\begin{multline}\label{Amp-inter-res3}
A_{emi}= \frac{1}{2}\text{i}\text{N}_1\text{N}_2\rho_1(2\pi)^2\pi^2\delta^2(\bar{k}^{(2)}-\bar{k}^{(1)})
\text{e}^{-\frac{1}{2}\pi\nu}\Big(\frac{k_0^{(2)}+k_1^{(2)}}{k_0^{(2)}-k_1^{(2)}}\Big)^{\frac{1}{2}\text{i}\nu}\\
\Big(\text{H}^{(1)}_{\text{i}\nu}(\text{i}\kappa \rho_1)\partial_{\rho}\text{H}^{(2)}_{\text{i}\nu}(\text{i}\kappa \rho_1)
- \partial_{\rho}\text{H}^{(1)}_{\text{i}\nu}(\text{i}\kappa \rho_1)\text{H}^{(2)}_{\text{i}\nu}(\text{i}\kappa \rho_1) \Big),
\end{multline}
where $\text{H}^{(a)}_{\text{i}\nu}(z)$, $a=1,2$, are the Hankel functions, and we have used the relation \eqref{KtoH}. Notice that the expression in the last parenthesis of \eqref{Amp-inter-res3} is the Wronskian involving the Hankel functions. It equals to
\be
-\frac{4\text{i}}{\pi\rho_1}.\label{Wronskian}
\ee
Plugging \eqref{Wronskian} in \eqref{Amp-inter-res3} we see that $\rho_1$ cancels out and after a few steps we get
\be\label{amemiss}
A_{emi}= \frac{\text{e}^{\pi\nu}}{(\text{e}^{2\pi\nu}-1)^{\frac{1}{2}}}\frac{1}{\sqrt{k^{\prime}_0}}
(\frac{k^{\prime}_0+k^{\prime}_1}{k^{\prime}_0-k^{\prime}_1})^{\frac{1}{2}\text{i}\nu}(2\pi)^2\delta^2({\bar{k}^{\prime}-\bar{k}}),
\ee
which numerically matches \eqref{ampli1}. Proceeding in the same way for the absorption amplitude we can obtain similar results.

Instead of explicitly compute the amplitudes, alternatively, one can figure out the thermal nature of the radiation by making on \eqref{compactA} a more elegant analysis, as presented in \cite{Hartle:1976tp} for a black hole and, more recently, in \cite{Padmanabhan:2019yyg} in the context of the Unruh effect. However, by proceeding with this analysis, we could miss some distinctive aspects of the non-thermal character of the particles detected by the Minkowski observer (the screen). In the subsequent discussion, we focus on these distinctive features of the radiation.

We have computed the emission amplitude using the flux of probabilities, and we have got the expected result, similar to \eqref{ampli1} and \eqref{ampli2}, and we can conclude that the radiation is thermal. So far, there are no new hints in this calculation leading to experimental confirmation of this process. Detecting the flux of matter particles radiated from the accelerated detector is almost impossible since the temperature is extremely low \eqref{temp} for the reachable accelerations in the lab. Although, we stress that now we have a better picture of how and where we could detect these radiated particles.

These amplitudes are independent of $R$ and $\rho_1$ as long as the screen is placed to the left of $\rho_1$.
Another experimental option we have would be to place the screen, as depicted in Fig. \ref{fig2}. In this case, there could be detection inside the right Rindler wedge and in both faces of the screen.

The amplitudes for this case have a different behavior; this is mainly because of now, the accelerated particle intersects the screen.
For instance, the left face perceives the process of emission and absorption of the accelerated detector happening in a finite time. It is worth emphasizing that, unlike for black holes where the spherical detector is placed outside the horizon, here we have the freedom of placing the screen in several positions relative to the accelerated detector. So, some of the results we present in the subsequent discussion are inherent to accelerated objects and do not have any analogous in black holes.\\

Let us sketch the calculation of the amplitude related to the experimental setup in Fig. \ref{fig2} . Suppose we prepare the screen in such a way there is no interference of the incoming waves with opposite momenta at the screen. We find it convenient to work under this assumption because we can treat left and right moving waves and amplitudes independently.

The amplitudes of detecting particles at the screen associated to the emission by a Rindler observer are
\bea\nonumber
A_{emi-} & = & \int\limits_{-\infty}^{\infty}dx^0\int\limits_{-\infty}^{\infty}d^2\bar{x}\ \text{J}_{1-}(x)\\
A_{emi+} & = & \int\limits_{-\infty}^{\infty}dx^0\int\limits_{-\infty}^{\infty}d^2\bar{x}\ \text{J}_{1+}(x)\nonumber
\eea
where the subscripts $\pm$ indicate the direction of the incoming wave Fig. \ref{fig2}. Under similar considerations as in our previous calculation, using \eqref{Amp-inter-res1}, we get\footnote{Note that the operator $\Big(\partial_{x^1}-\text{i}k_1^{(2)} \Big)$ acting on the function $\int\limits_{-\infty}^{\infty}d\tau \text{exp}\Big(-\text{i}|x^1-\rho_1\ \text{cosh}(\tau)||k_1^{(2)}|\Big)$ selects the $\tau$ intervals according to the sign of $\Big(x^1-\rho_1\ \text{cosh}(\tau)\Big)$, and the sign of $k_1^{(2)}$.}
\begin{multline}\label{Amp-inter-res_m}
A_{emi-}= \text{i}\text{N}_1\text{N}_2\rho_1(2\pi)^2\delta^2(\bar{k}^{(2)}-\bar{k}^{(1)})
\Big(\text{K}_{\text{i}\nu}(\kappa \rho_1)\partial_{\rho}- \partial_{\rho}\text{K}_{\text{i}\nu}(\kappa \rho_1) \Big)\\
\Big(\int\limits_{-\infty}^{-\tau_R}d\tau+\int\limits_{\tau_R}^{\infty}d\tau \Big) \text{e}^{\text{i}k_1^{(2)}y^1+\text{i}k_0^{(2)}y^0-\text{i}\nu\tau}\Big|_{ \rho=\rho_1},
\end{multline}
with $k_1^{(2)}<0$,
and
\begin{multline}\label{Amp-inter-res_p}
A_{emi+}=-\text{i}\text{N}_1\text{N}_2\rho_1(2\pi)^2\delta^2(\bar{k}^{(2)}-\bar{k}^{(1)})
\Big(\text{K}_{\text{i}\nu}(\kappa \rho_1)\partial_{\rho}- \partial_{\rho}\text{K}_{\text{i}\nu}(\kappa \rho_1) \Big)\\
\int\limits_{-\tau_R}^{\tau_R}d\tau \text{e}^{\text{i}k_1^{(2)}y^1+\text{i}k_0^{(2)}y^0-\text{i}\nu\tau}\Big|_{ \rho=\rho_1},
\end{multline}
with $k_1^{(2)}>0$,
where $\tau_R$, is the time where the accelerated particle meets the screen, $\tau_R=\text{arccosh}(\frac{R}{\rho_1})$. Notice that none combination of the integrals in \eqref{Amp-inter-res_m} and \eqref{Amp-inter-res_p} reproduces \eqref{Amp-inter-res2}. We also have similar results for the absorption amplitudes.

From this result, we can conclude that when the screen is placed to the right of $x^1=\rho_1$, Fig. \ref{fig2} we can find deviation from thermality. {\it From the Minkowski perspective, the thermal character of the radiation emitted by the accelerated object depends upon where this radiation is collected}. It is the reason why we mentioned in the introduction that, strictly speaking, it could be called ``a non-thermal detection.'' It is a new result that can not be derived from the direct calculation of the amplitude as in \eqref{ampli1} and \eqref{ampli2}, and can not be found in any of the detector models in the literature, see \cite{Sriramkumar:1999nw} and references therein for a discussion related to detector models.

We would like to stress that we are discussing the detection of matter particles emitted by the accelerated detector, but the previous results also hold for electromagnetic radiation. To our knowledge, this kind of experimental setups, and results have not been presented before in the literature.

Let us now briefly present the third experimental option, the screen placed at a very large positive $R$, or for mathematical purposes $R\rightarrow \infty$. This limit can be taken on expressions \eqref{Amp-inter-res_m} and \eqref{Amp-inter-res_p}. When $R\rightarrow \infty$, $\tau_R\rightarrow \infty$. The two integrals of \eqref{Amp-inter-res_m} vanish, while \eqref{Amp-inter-res_p} coincides with \eqref{Amp-inter-res2} and hence \eqref{amemiss}. So, when the screen is placed at $R\rightarrow \infty$ we recover the thermal behaviour of the radiation.

One crucial feature of \eqref{Amp-inter-res2} is that the integral involved needs regularization, see Appendix \ref{appen:B}. However, the integral
\eqref{Amp-inter-res_p} can be highly oscillating, but for finite $\tau_R$, it is finite. Taking this into consideration in the next subsection, we shall numerically explore the amplitudes $A_{emi+}$, and $A_{abs+}$, and their associated probabilities.
Could this deviation from thermality open a window for detecting more easily the radiated particles? The answer to the previous question can be found in the next section.

\subsection{\label{subsec:3.2} Non-thermal Radiance}

The amplitude of detecting particles at the screen related to $\text{J}_{1+}(x)$, Fig. \ref{fig2}, is given by \eqref{Amp-inter-res_p}. Similarly for $A_{abs+}$. The probability of detection follows the same rules as in \eqref{Total_Proba}.

\begin{figure}[]
\centering
\includegraphics[width=.8\textwidth]{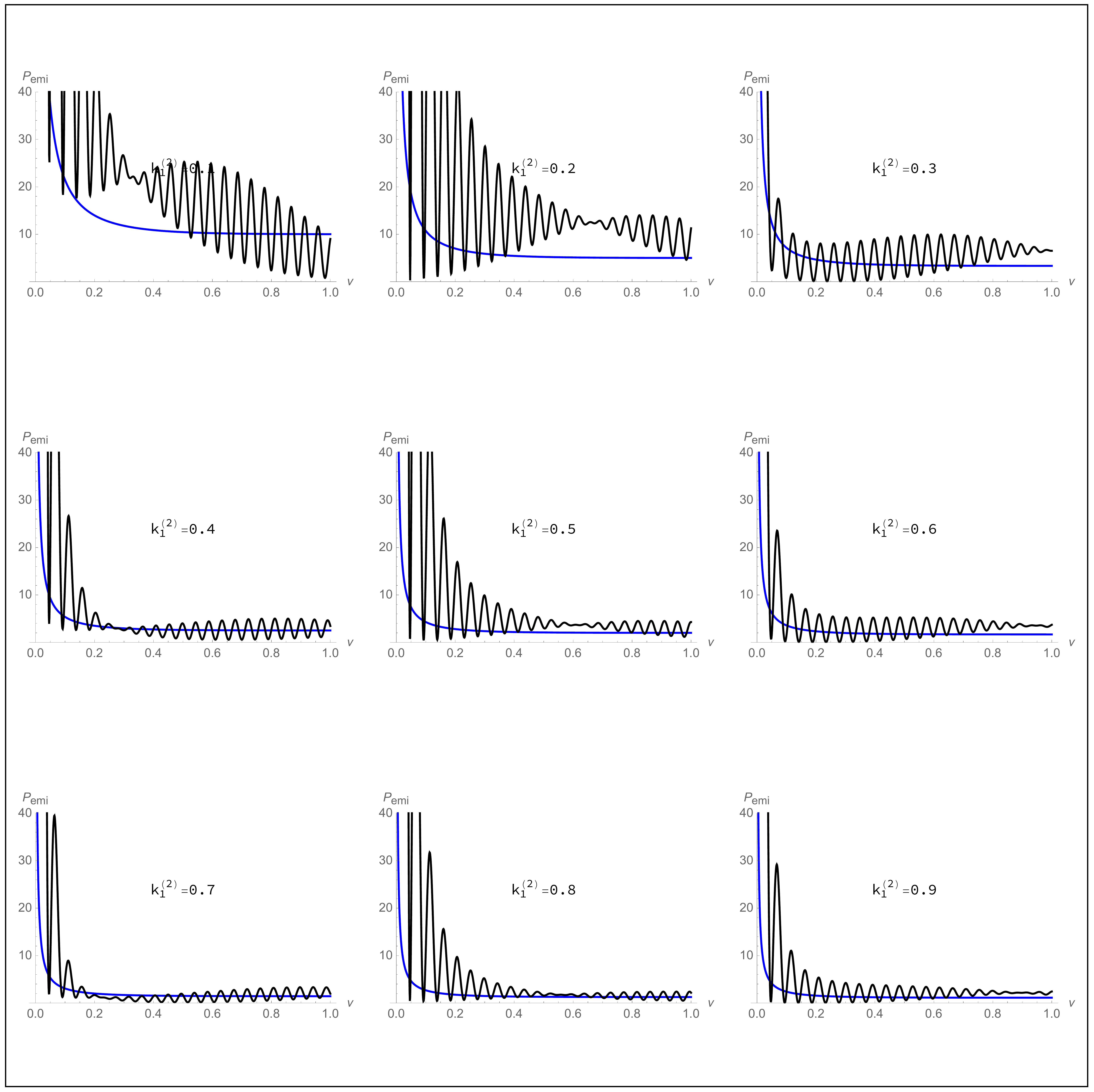}
\caption{\sl Probability of detection at the screen associated to the emission by the accelerated object. In blue the thermal probability. In black the non thermal probability for $R=10$ and $\rho_1=1$.}\label{fig3}
\end{figure}

\begin{figure}[]
\centering
\includegraphics[width=.8\textwidth]{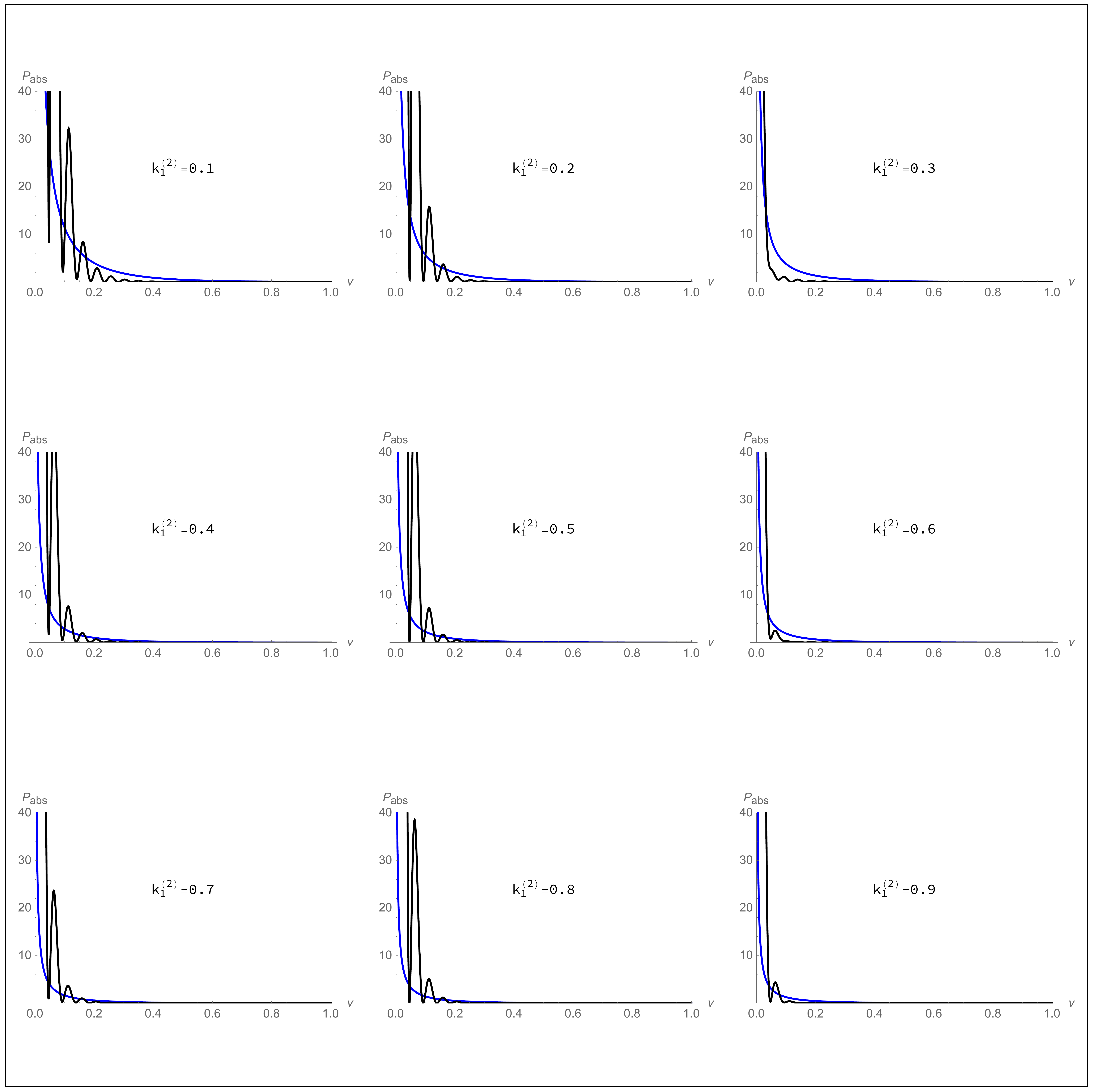}
\caption{\sl Probability of detection at the screen associated to the absorption by the accelerated object. In blue the thermal probability. In black the non thermal probability for $R=10$ and $\rho_1=1$.}\label{fig4}
\end{figure}

Lest us first present the plot for different values of $k_1^{(2)}$, with $k_2^{(2)}=k_3^{(2)}=0$, of the probability of detection from the Minkowski perspective, i.e., the screen, associated to the emission and absorption of a Rindler mode by the accelerated detector Fig. \ref{fig3} and Fig. \ref{fig4}. The probability takes its maximum value when $k_2^{(2)}=k_3^{(2)}=0$. For $k_2^{(2)}\neq0$, and $k_3^{(2)}\neq0$, the amplitude rapidly falls to zero.

From Fig. \ref{fig3} and Fig. \ref{fig4} we can see the oscillatory behavior of the non-thermal probabilities. We have used natural units $\hbar =\text{c}= k_B=\bold{a}=1$. By restoring the constants, we see that the probability is still insignificant. What is remarkable is that for certain values of the frequency $\nu$, the non-thermal probability of detection is greater than the thermal. Perhaps in this setup, we could enhance the probability of detection by considering several accelerated particles, i.e., several Rindler and Minkowski modes in the initial and final state, respectively.

We are referring to this result as ``non-thermal''; however, so far, we do not know whether it is associated with a thermal process or not. We only know that the amplitudes do not correspond to those associated with a thermal process. An excellent test to diagnose the thermal nature of a given process comes from the ratio $\frac{P_{abs}}{P_{emi}}$.

\begin{figure}[]
\centering
\includegraphics[width=.8\textwidth]{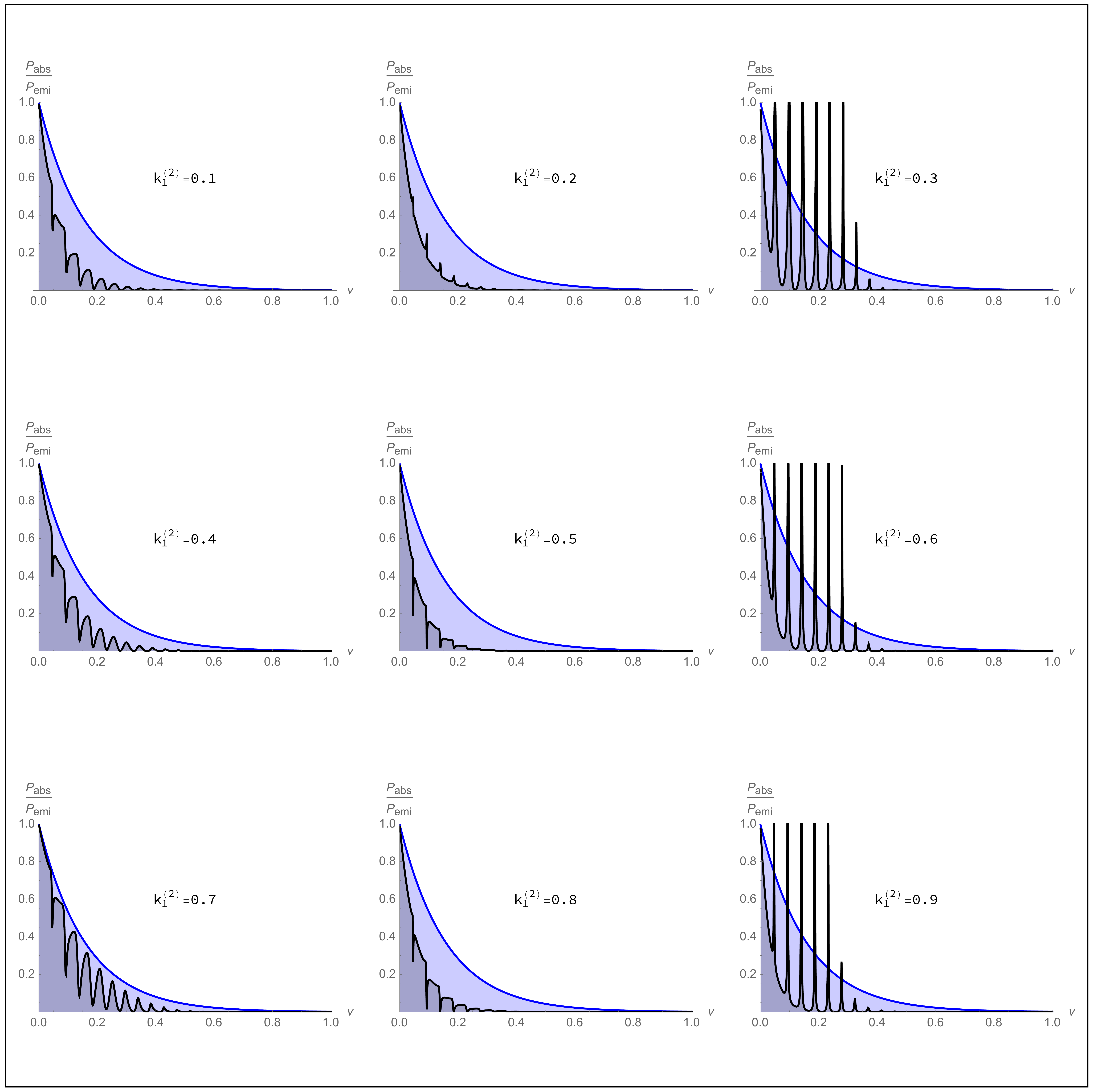}
\caption{\sl Ratio between the probabilities associated to emission and absorption by the accelerated object. In blue the thermal ratio. In black the non thermal ratio for $R=10$ and $\rho_1=1$.}\label{fig5}
\end{figure}

In Fig. \ref{fig5} we present a comparison between the ratio $\frac{P_{abs}}{P_{emi}}=\text{e}^{-2\pi\nu}$ for a thermal detection and the same ratio but related only to $\text{J}_{1+}(x)$.

From Fig. \ref{fig5} we can see that from the Minkowski perspective, under the conditions of the experiment, the process looks completely non-thermal. In this case, we can not associate a temperature to this radiation.

\begin{figure}[]
\centering
\includegraphics[width=.6\textwidth]{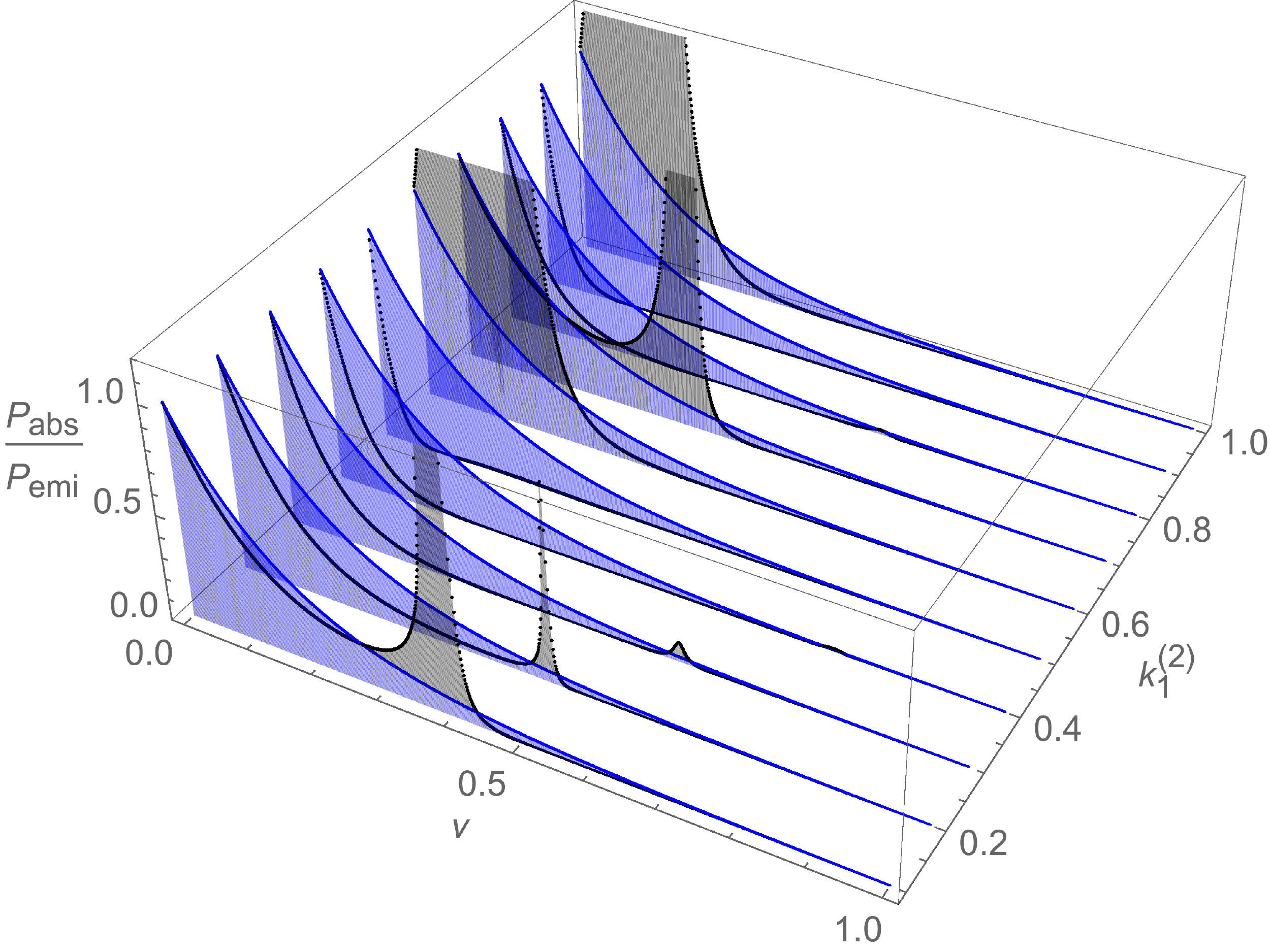}
\caption{\sl Ratio between the probabilities associated to emission and absorption. In blue the thermal ratio. In black the non thermal ratio. $R=10$, $\rho_1=1$ and $(k_2^{(2)})^2+(k_3^{(2)})^2=0.6$\ . }\label{fig6}
\end{figure}

For the sake of completeness we present the plot of the ratio between the probabilities when $k_2^{(2)}\neq0$, and $k_3^{(2)}\neq0$ in Fig. \ref{fig6}. We do not present the probabilities associated to the emission and absorption independently because they are difficult to appreciate in the figure.

From Fig. \ref{fig6} we can see that for the intervals, we are considering for the momentum $k_1^{(2)}$ and the frequency $\nu$; there are values where we can find large deviations from the thermal behavior.

\section{Conclusions}

In these notes, we have proposed a new phenomenon that we have called {\it Rindler observer sublimation}.
We have followed the logical arguments in reference \cite{Hartle:1976tp}, where similar calculations have been presented by Hartle and Hawking for a black hole, to conclude that this process is possible. The critical assumption here is that an accelerated detector made of given particles might probe the quantum vacuum of the same kind of particles the detector is made.

 To make the presentation more pedagogical, in section \ref{sec:2}, we reviewed the quantization of a massive scalar field in Rindler space, highlighting that different boundary conditions can be used. This section is complemented with appendix \ref{appen:A}, where the ordinary quantization of the massive scalar field has been presented, and can be contrasted with section \ref{sec:2}. In section \ref{sec:3}, we computed in two different ways the amplitudes of emission and absorption of an accelerated detector. We have made a parallel discussion between our calculation and reference \cite{Hartle:1976tp}, where a similar calculation has been presented for a black hole. Our calculation shows that in the context of accelerated detectors, one gets similar results to the amplitudes one gets for a black hole. Three experimental setup has been presented. In particular, one of them shows that under certain circumstances, one can get a non-thermal outcome. The non-thermal probability has been numerically explored. The graphics of the comparison between the thermal and no-thermal amplitudes has been presented at the end of section \ref{sec:3}.

Finding that the amplitude \eqref{compactA} is non-vanishing for the process we are considering is an irrefutable evidence that a Rindler observer may sublimate (or evaporate).
The calculation of the emission and absorption amplitudes from the Minkowski perspective in section \ref {sec:3} leads us to conclude that a screen placed in the neighborhood of a uniformly accelerated particle could detect some thermal radiation made of matter particles. Besides, we have found that when the uniformly accelerated particle intersects the screen, the collected radiation does not have a thermal distribution.This new result does not have analogous in black holes.

 We find in this deviation from thermality a small window for confirmation of this process. Also, if confirmation is achieved, it would shed some light on the Unruh effect. The fact that we can not associate a temperature to the radiation in this particular case, and that for some values of the frequency $\nu$ the flux of detected particles could be greater than the flux when the radiation is thermal Fig. \ref{fig3} and Fig. \ref{fig4}, make confirmation plausible.

We have presented two different ways for computing the amplitude, \eqref{ampli1} and \eqref{ampli2} and, \eqref{H-ampli} (or \eqref{compactA}). Although they are quantitatively equal, we have to stress that they also are qualitatively different. The interpretation is clear, \eqref{ampli1}, and \eqref{ampli2} gives us the transition amplitude between two states defined over two spacelike surfaces. While \eqref{H-ampli} (or \eqref{compactA}) gives us the amplitude of detection at some particle detector, in our case, the screen. Note that the integrals \eqref{H-ampli} (or \eqref{compactA}) are taken over timelike surfaces.

In this work, for simplicity, we did not consider the backreaction due to sublimation. To our purpose, proving that the process involved has a non-vanishing amplitude was sufficient.
We have implicitly assumed that the accelerated detector is heavy enough.  In this way, losing a few electrons does not affect its trajectory.

The kind of accelerated detector we have presented in this work is that accelerated box with particles within it we mentioned in the introduction. The details of the confining potential are not relevant in the calculation presented here. They would show up only in an overall factor in front of the amplitudes, and for simplicity, we have decided to omit it.

The next step in this project could be to engineer a detector model that takes into account the backreaction. This way, we could make more precise statements at any stage of the sublimation. Namely, we could track at any time the whole process in more realistic systems. In this model, the details of the confining potential would be relevant.

We want to stress one more time that we have focused on the sublimation process for matter particles, but all the formalism presented here applies to electromagnetic radiation too. Of course, for electromagnetic radiation does not make any sense to talk about sublimation. In particular, the results presented in section \ref {subsec:3.2} regarding the non-thermal character of the radiation works for (massless scalar fields) electromagnetic fields as well. So instead of designing an experiment to confirm the sublimation of an accelerated object, one could design an experiment to confirm the results of section \ref {subsec:3.2} related with the non-thermal character for the electromagnetic radiation.

Although we have presented some evidence that the sublimation of Rindler observers is possible, this is a minuscule effect, and like for black hole evaporation, much more study is needed before establishing that it can occur in nature.

\section*{Acknowledgments}

We are grateful to H. Casini, Pablo Diaz, and Kanghoon Lee for discussions and useful comments. We would especially like to thank Pablo Diaz, who participated in the first steps of this research.

\appendix

\section{\label{appen:A} Bogoliubov Coefficients, Canonical Quantization }

In the canonical quantization in Rindler space instead of imposing the boundary condition \eqref{B.C}, we impose.
\bea\label{B.C.c.q}
\varphi_R(\tau,\rho,\bar{x}) & = & \varphi_M(x^0(\tau,\rho),x^1(\tau,\rho),\bar{x}), \nonumber\\
\partial_\tau\varphi_R(\tau,\rho,\bar{x}) & = & \partial_\tau\varphi_M(x^0(\tau,\rho),x^1(\tau,\rho),\bar{x}),
\eea
at some initial Rindler time $\tau$.

In this case we have the equations
\be
\text{e}^{-\text{i}\nu \tau}b_{(\nu,-\bar{k})}+\text{e}^{\text{i}\nu \tau}b^{\dagger}_{(\nu,\bar{k})}=
(2\nu)^{\frac{1}{2}}\int\limits_{0}^{\infty}d\rho\int\limits_{-\infty}^{\infty}d^2x\frac{\psi_{\nu,\kappa}(\rho)}{\rho}\text{e}^{-\text{i}\bar{k}\cdot\bar{x}}\varphi_{M}(\tau,\rho,\bar{x}),
\ee
and
\be
\text{e}^{-\text{i}\nu \tau}b_{(\nu,-\bar{k})}-\text{e}^{\text{i}\nu \tau}b^{\dagger}_{(\nu,\bar{k})}= \\
\text{i}\frac{(2\nu)^{\frac{1}{2}}}{\nu}\int\limits_{0}^{\infty}d\rho\int\limits_{-\infty}^{\infty}d^2x\frac{\psi_{\nu,\kappa}(\rho)}{\rho}\text{e}^{-\text{i}\bar{k}\cdot\bar{x}}\partial_\tau\varphi_{M}(\tau,\rho,\bar{x}).
\ee
Solving for $b_{(\nu,\bar{k})}$, we get
\be\label{b-bogo-cano}
b_{(\nu,\bar{k})}=\frac{1}{2}\text{e}^{\text{i}\nu \tau}(2\nu)^{\frac{1}{2}}(1+\frac{\text{i}}{\nu}\partial_\tau)\\ \int\limits_{0}^{\infty}d\rho\int\limits_{-\infty}^{\infty}d^2x\frac{\psi_{\nu,\kappa}(\rho)}{\rho}\text{e}^{\text{i}\bar{k}\cdot\bar{x}}\varphi_{M}(\tau,\rho,\bar{x}).
\ee

Now we can use the result \eqref{Int.coeff} in equation \eqref{b-bogo-cano} to finally obtain \eqref{pre-Bogo}. We stress that the boundary conditions in the canonical quantization are different to the ones used in section \ref{sec:2}, but certainly they are equivalent. Using \eqref{rela1} and \eqref{rela2}, \eqref{pre-Bogo} can be written as
\be\label{Bogoliubov coefficients}
b_{(\nu,\bar{k})}=\frac{1}{2\pi}\frac{1}{(\text{e}^{2\pi\nu}-1)^{\frac{1}{2}}}\int\limits_{-\infty}^{\infty}\frac{dp_1}{\sqrt{p_0}}(\frac{p_0+p1}{p_0-p_1})^{-\frac{1}{2}\text{i}\nu}\Big(\text{e}^{\pi\nu} a_{(p_1,\bar{k})}+ a^{\dagger}_{(p_1,-\bar{k})}\Big),
\ee
where $p_0=\sqrt{p_1^2+\kappa^2}$, \eqref{kappa}.

\section{\label{appen:B} Integral Representation of the Hankel Functions}

The integral representation of the Hankel functions can be found in \cite{DLMF}. Here we present the integral representation of $H^{(2)}_{\nu}\big((z^2-\zeta^2)^{\frac{1}{2}}\big)$, in connection with the Amplitude calculation and the wave function in the Rindler wedge.

This function can be represented as
\be
H^{(2)}_{\nu}\big((z^2-\zeta^2)^{\frac{1}{2}}\big)=-\frac{1}{\pi\text{i}}\text{e}^{\frac{1}{2}\nu\pi\text{i}}
\Big(\frac{z+\zeta}{z-\zeta}\Big)^{-\frac{1}{2}\nu}
\int\limits_{-\infty}^{\infty}d\tau \ \text{e}^{-\text{i}z \text{cosh}(\tau)-\text{i}\zeta \text{sinh}(\tau)-\nu\tau},
\ee
where $\nu, z,\zeta \in\mathbb{C}$, and $\text{Im}(z\pm\zeta)<0$.

The integral in \eqref{Amp-inter-res2} can be rewritten as
\be\label{analytical_int}
\int\limits_{-\infty}^{\infty}d\tau \ \text{e}^{-\text{i}(-k_1^{(2)}\rho)\text{cosh}(\tau)-\text{i}(-k_0^{(2)}\rho)\text{sinh}(\tau)-\text{i}\nu\tau}.
\ee
Now, the analytical extension of \eqref{analytical_int} is defined as
\be
(-\text{i}\pi)\text{e}^{\frac{1}{2}\nu\pi}\Big(\frac{k_1^{(2)}+k_0^{(2)}}{k_1^{(2)}-k_0^{(2)}}\Big)^{\frac{1}{2}\text{i}\nu}
H^{(2)}_{\text{i}\nu}\Big(\rho\sqrt{(k_1^{(2)})^2-(k_0^{(2)})^2}\Big).
\ee
The on shell condition $(k_0^{(2)})^2-(k_1^{(2)})^2-\kappa^2=0$ reduces the integral \eqref{analytical_int} to
\be
(-\text{i}\pi)\text{e}^{\frac{1}{2}\nu\pi}\Big(\frac{k_1^{(2)}+k_0^{(2)}}{k_1^{(2)}-k_0^{(2)}}\Big)^{\frac{1}{2}\text{i}\nu}
H^{(2)}_{\text{i}\nu}(\text{i}\kappa\rho),
\ee
which is the result we have used in \eqref{Amp-inter-res3}.

We can also use the integral representation of the Hankel function to relate the one particle wave functions \eqref{emiWF} and \eqref{absWF} in Rindler space with a wave package in Minkowski space. They are related as: for the emission wave function
\be\label{wfRM1}
\text{e}^{-\text{i}\nu\tau}\text{K}_{\text{i}\nu}(\kappa\rho)=\frac{1}{2}\text{e}^{\frac{1}{2}\nu\pi}
\int\limits_{-\infty}^{\infty}\frac{dk_1}{k_0}\Big(\frac{k_0+k_1}{k_0-k_1}\Big)^{\frac{1}{2}\text{i}\nu}\text{e}^{-\text{i}(k_0x^0+k_1x^1)},
\ee
for the absorption wave function
\be\label{wfRM2}
\text{e}^{\text{i}\nu\tau}\text{K}_{\text{i}\nu}(\kappa\rho)=\frac{1}{2}\text{e}^{-\frac{1}{2}\nu\pi}
\int\limits_{-\infty}^{\infty}\frac{dk_1}{k_0}\Big(\frac{k_0+k_1}{k_0-k_1}\Big)^{-\frac{1}{2}\text{i}\nu}\text{e}^{\text{i}(k_0x^0+k_1x^1)}.
\ee

Here we have followed three steps. First, we have performed the change of variables
\bea \nonumber
\tau & = & \frac{1}{2}\text{Log}\Big(\frac{x^1+x^0}{x^1-x^0} \Big), \nonumber\\
\rho & = &\sqrt{(x^1)^2-(x^0)^2}, \nonumber
\eea
which is the inverse transformation of \eqref{transf0}.
Second, we have used the relation
\be
\text{K}_{\text{i}\nu}(z)=\frac{\pi}{2}(\text{i})^{\text{i}\nu+1}\text{H}^{(1)}_{\text{i}\nu}(\text{i}z)\label{KtoH},
\ee
and the integral representation of $\text{H}^{(1)}_{\text{i}\nu}(\text{i}z)$, see \cite{DLMF}.
Finally, the change of variables
\bea \nonumber
k_0 & = & \kappa\ \text{cosh}(t), \nonumber\\
k_1 & = & \kappa\ \text{sinh}(t), \nonumber
\eea
brings the wave functions to the desired form.

It is worth to emphasize that \eqref{wfRM1} and \eqref{wfRM2}, are valid only on the overlap between Rindler and Minkowski space. With this, we can conclude that the Rindler one-particle wave function can be seen as a fully localized (inside the Rindler wedge) wave package from the Minkowski perspective.

\end{document}